\documentclass{PoS}

\usepackage{amsmath}
\usepackage{placeins}

\title{Improved Sampling Algorithms for Lattice QCD}

\ShortTitle{Algorithms in Lattice QCD}

\author{\speaker{Arjun Singh Gambhir}\\
        Department of Physics, The College of William \& Mary, Williamsburg, VA 23185, U.S.A\\
        E-mail: \email{asgambhir@email.wm.edu}}

\author{Kostas Orginos\\
        Department of Physics, The College of William \& Mary, Williamsburg, VA 23185, U.S.A.\\
        Thomas Jefferson National Accelerator Facility, Newport News, VA 23606, U.S.A\\
        E-mail: \email{kostas@jlab.org}}

\abstract{Reverse Monte Carlo (RMC) is an algorithm that incorporates stochastic modification of the action as part of the process that
 updates the fields in a Monte Carlo simulation.  Such update moves have the potential of lowering or eliminating potential barriers that
 may cause inefficiencies in exploring the field configuration space. The highly successful Cluster algorithms for spin systems can be derived from the RMC framework. In this work we apply RMC ideas to pure gauge theory, aiming to alleviate the critical slowing down observed in the topological charge evolution as well as other long distance observables. 
 We  present various formulations of the basic idea and report on our numerical experiments with these algorithms.}

\FullConference{The 32nd International Symposium on Lattice Field Theory,\\
		23-28 June, 2014\\
		Columbia University New York, NY}

\begin{document}

\section{Introduction}
Critical slowing down is one of the main hindrances to reliable computation with Monte Carlo simulations in Lattice Field Theory.
  This is the phenomenon of growing computational costs when approaching a theory's critical point because of large autocorrelation times. The severity of this slow down is dependent on the algorithm used and the observable of interest. In order to obtain reliable statistics, all Monte Carlo simulations must have a much longer simulation time than the autocorrelation time of the measured observable.

In Lattice QCD, when approaching the continuum limit, the topological charge develops a very large autocorrelation time~ \cite{Sommer}. This is because of the large barriers between topological sectors that develop at small lattice spacing, thereby reducing the tunneling rate between these sectors. Large autocorrelation times for the topological charge as well as for other long distance observables, represent a major challenge for current QCD simulations. 

Recently, open boundary conditions in the Euclidean time direction instead of periodic have been proposed~\cite{open}, as a solution to this problem. This allows topological charge to flow through the boundaries of the lattice and alleviate critical slowing down. In recent work the effectiveness of open boundary conditions has been studied extensively with Hybrid Monte Carlo (HMC) simulations and flow of topological charge has been shown to be a diffusive process \cite{Greg}.

Although open boundary conditions decrease the autocorrelation time at fine lattice spacing, translation invariance in the time direction is lost in these simulations. In certain cases it is preferable to use periodic boundaries. This gives motivation for new algorithms that decrease autocorrelation time yet retain periodic boundary conditions to be explored. In this work we explore possible modifications to the Monte Carlo update process that may reduce autocorrelation times. One such algorithm is Reverse Monte Carlo (RMC)~ \cite{Mak} that was found to be very effective in scalar field theories in two dimensions.

\section{Reverse Monte Carlo}
Reverse Monte Carlo (RMC) is an algorithm that apart from updating the fields of a theory, also introduces local stochastic modification of the action itself. 
We write the action for the system we are simulating as 
\begin{equation}
S = \sum_n{{\cal L}(n)} \,,
\end{equation}
where $n$ is an index of the lattice site and ${\cal L}(n)$ is the local action density.
We introduce local modifications of the action density ${\cal L}_s(n)$ that can occur with probability $P(n)$ on each site $n$ given by
\begin{equation}
P(n)  =  \frac{ e^{\Delta {\cal L}(n)} }{C} =   \frac{ e^{{\cal L}(n) - {\cal L}_s(n)} }{C} \,,
\end{equation}
where $C$ is a normalization constant that ensures the switch probability is $P(n)\le 1$ for all possible field values on site $n$. One such convenient choice is
\begin{equation}
C  = e^{max(\Delta {\cal L}(n))}\,,
\end{equation}
where the  maximum is taken over all possible values of the fields in the action.
Note that the choice of $C$ is not unique and any choice that ensures $P(n)\le 1$ is sufficient.
On a given field configuration during the simulation, local changes of the action are proposed and accepted with probability $P(n)$. 
However, in order to have a Monte Carlo update that converges to the desired probability distribution dictated by the original action, the action density of sites that do not switch needs to also be modified as shown in~\cite{Mak}. The resulting action density
after switching becomes
\begin{equation}
{\cal L}'(n) = \left\{
\begin{array}{cr}
{\cal L}_s(n) & \text{sites that switched}\\
{\cal L}(n) - \operatorname{ln}(1-P(n)) & \;\;\;\; \text{sites that did not switch}
\end{array}
\right.\,.
\end{equation}
Subsequently the fields are updated, but now using the new modified action for the system. Any update method of the fields can be incorporated, however maximal efficiency can be achieved if specific features of the modified action are exploited as was done in~\cite{Mak}. For details on the proof of validity of this method the reader is referred to~\cite{Mak}. It turns out that this approach represents a generalization to cluster algorithms for spin systems and in fact the original Swendsen-Wang cluster algorithm \cite{Swendsen} for the Ising model can be derived in this framework.  

\section{Application of RMC to QCD}
   The application of RMC to Lattice QCD is straight forward. Our example is with pure gauge theory and the Wilson gauge action. Incorporation of fermions is straight forward given that our basic field update algorithm is Hybrid Monte Carlo (HMC) and thus it is omitted here. Furthermore, the issue of large autocorrelations of the topological charge arises in pure gauge theory as well.
 
 The Wilson gauge action is written as
 \begin{equation}
S_G\left[U\right]=\beta\ \sum_{n,\mu,\nu}\frac{1}{N_c}\operatorname{ReTr}\left[\textbf{1}-U_{\mu\nu}(n)\right]=
\beta\ \sum_{n,\mu,\nu} A_{\mu\nu}(n)\,,
\end{equation}
where $U_{\mu\nu}(n)$ is the plaquette and $A_{\mu\nu}(n)$ is the action density on site $n$. 
The pure gauge partition function is
\begin{equation}
{\cal Z}=\int D[U] \ e^{-S_G[U]}\,,
\end{equation} 

The local modification to the action density we are proposing is to introduce a plaquette and site dependent coupling $\beta_{\mu\nu}(n)$. The coupling constant can be allowed to fluctuate between the original value $\beta$ and a lower value $\alpha$. If an appreciable fraction of plaquettes acquire a small coupling constant $\alpha$, one can hope
that updates of the gauge fields with this modified action can allow topological charge fluctuations that are suppressed with the original action. 

The probability for a plaquette to switch is dependent on the difference of unswitched and switched terms  
\begin{equation}
P_{n,\mu,\nu}=Ce^{\left(\beta-\alpha)\right)A_{\mu\nu}(n)}\,,
\end{equation}
with
 \begin{equation}
C  = e^{max[(\beta - \alpha) A_{\mu\nu}(n)]}\,.
\end{equation}
Due to SU(3) being a compact group, the trace of the plaquette is bounded, hence there exists a unique value for C.

The new action to be used in the subsequent update is 
\begin{equation}
S_G'\left[U\right]=\sum_{n,\mu,\nu} \ \tilde{\beta}_{\mu\nu}(n) \ A_{\mu\nu}(n)
- \sum_{n\in {\cal N},\rho,\sigma} \  {\rm ln}(1-P_{\rho,\sigma}(n))
\end{equation}
where $\cal N$ is the set of sites which did not switch, and retained the original coupling $\beta$.
In our application we use HMC for the subsequent field update, hence the inclusion of fermions is straight forward.

\subsection{Continuous Beta}
Direct application of RMC to QCD results in small switch rates, especially if a large change in $\beta$ is allowed. 
For this reason one may want to modify the algorithm in a way that gives a range of $\beta$ values. Following this idea, we introduce a new method that allows for a continuous variation of the coupling constant in the interval $[\alpha, \beta]$. In this case the switch probability is given by:
\begin{eqnarray}
p(\tilde{\beta}_{\mu\nu}(n)) \ \mathrm{d}\tilde{\beta}_{\mu\nu}(n)=
\mathrm{d}\tilde{\beta}_{\mu\nu}(n)\ e^{\left(\beta-\tilde{\beta}_{\mu\nu}(n)\right)A_{\mu\nu}(n)}\frac{A_{\mu\nu}(n)}{e^{\left(\beta-\alpha\right)A_{\mu\nu}(n)}-1}\,,
\end{eqnarray}
resulting in a new action with a new field $\tilde \beta_{\mu\nu}$ to be simulated given by
\begin{eqnarray}
S_G'\left[U,\tilde\beta_{\mu\nu}\right]= \sum_{n,\mu,\nu} \tilde{\beta}_{\mu\nu}(n) A_{\mu\nu}(n)
-{\rm ln}\left\{\frac{1}{A_{\mu\nu}(n)}\left(e^{\left(\beta-\alpha\right)A_{\mu\nu}(n)}-1\right)\right\}.
\label{eq:modS_Cbeta}
\end{eqnarray}
 We call this algorithm ``Continuous Beta''. Unlike traditional RMC, Continuous Beta gives a log term for every plaquette. 
The advantage of this method is that $\alpha$ can be set to 0 and plaquettes are more likely to switch close to this lower bound. The logarithmic piece that accompanies each interaction must be accounted for however, and in general introduces nontrivial tunneling properties. Note that in this case the chosen probability distribution for the coupling constant integrates to one i.e.
\begin{equation}
\int_\alpha^\beta  p(\tilde{\beta}_{\mu\nu}(n)) \ \mathrm{d}\tilde{\beta}_{\mu\nu}(n) = 1\,.
\label{eq:normCbeta}
\end{equation}

There are other possibilities for implementing a continuous coupling algorithm, which are currently under investigation.
The validity of this algorithm will be explained in the next section where an alternative proof for the RMC algorithm is also presented.

\section{Path Integral Interpretation}

The authors in \cite{Mak} proved that Reverse Monte Carlo respects detailed balance of the original theory. When applying RMC to Lattice Field Theory, the same conclusion can be reached using a simple argument based on the path integral that defines the correlation functions to be simulated. 
The fluctuating coupling constants  $\tilde{\beta}_{\mu\nu}$ can be thought of as a new site dependent field. The action of this new field has been constructed in such way that after integrating over this new field the original path integral with fixed coupling constant is recovered. Therefore, simulating the new augmented theory results in the same correlation functions as the original theory. To be more precise, using Eq.~\ref{eq:normCbeta} in the case of the Continuous Beta algorithm, the following identity holds for the partition function
\begin{equation}
{\cal Z}=\int D[U] \ e^{-S_G[U]} = \int D[U] \ e^{-S_G[U]} \int_\alpha^\beta \prod_n p(\tilde{\beta}_{\mu\nu}(n)) \ \mathrm{d}\tilde{\beta}_{\mu\nu}(n)
\end{equation}
which results in
\begin{equation}
{\cal Z}=\int D[U] D[\tilde \beta_{\mu\nu}] \ e^{-S_G'[U,\tilde \beta_{\mu\nu}]}.
\end{equation}
Hence, correlation functions of the original theory can be written as
\begin{equation}
\langle {\cal O}\rangle = \frac{1}{\cal Z} \int D[U] D[\tilde \beta_{\mu\nu}]  \ {\cal O}(U) \ e^{-S_G'[U,\tilde \beta_{\mu\nu}]}\,,
\end{equation}
where $ {\cal O}(U)$ is the desired correlation function of gauge fields. 
From these identities it is clear that the proposed update algorithm is nothing  but a normal update algorithm for the new augmented theory that first updates the coupling constant field and subsequently updates the gauge fields. 

For RMC the same formalism holds with the distinction that the integral over the fluctuating coupling constant has to be
replaced with a sum over two distinct values of $\beta$. In this case the sum of probabilities for having the coupling constant to be $\beta$ and probability of switching to $\alpha$ add up to one. Hence the manipulations done for the Continuous Beta case also hold. Furthermore, this point of view gives an easy way to generalize RMC to the case of multiple distinct values for the coupling constant. In addition, the path integral formalism makes it easy to design other update algorithms with fluctuating terms in the action. 

Now that we have established the correctness of this class of algorithms (RMC and the Continuous Beta generalization),
we may attempt to tune the update algorithm of the augmented action. One idea we tried was to introduce a freezing parameter 
into the update step of the coupling constants. Instead of switching the coupling on every update, we froze it for $\kappa$ steps. Setting a moderately high $\kappa$ ensures that the simulation has enough time to take advantage of potentially lowered barriers between topological sectors. From the work done in \cite{Greg}, it is observed that the topological charge fluctuations obey a simple diffusion and tunneling model. Our action modifications may enhance the tunneling rates locally, however enough simulation time needs to be given for the diffusion process to propagate topological fluctuations. 
Careful study of the dependence of the algorithm on $\kappa$ is needed in order to construct an efficient algorithm. 


\section{Simulation Results}

We performed numerical experiments  on a $32^4$ lattice with $\beta=6.179$ with the pure gauge Wilson action. The simulation algorithm of choice was HMC with a fourth order integrator. The integrator parameters used resulted in an acceptance rate of 
$\approx 90\%$.

The Continuous Beta simulations utilized a lower bound for the coupling constant  of $\alpha=0$ and $\alpha=5$. The basic RMC algorithm was run with a modified coupling of $\alpha=5$. Additionally, separate experiments with RMC $\alpha=5$ that froze the stochastic modification for 100 or 200 steps were also conducted. We computed expectation values and autocorrelation times for the plaquette and smeared plaquette, and topological charge using the above simulation algorithms as well as 
the basic HMC algorithm. The pure HMC simulation is marked the "Base" case in subsequent tables. The smearing method used to compute topological charge and the smeared plaquette was gradient flow \cite{Gradient} with a flow time of $\tau=2$ in lattice units.

\begin{table}[!ht]
\caption{Plaquette 10,000 Trajectories} 
{\centering  
\begin{tabular}{c c c} 
\hline\hline 
Algorithm & Value\ & Integrated Autocorrelation \\ [0.5ex] 
\hline 
Base & 0.6116998(3) & 3.26(3) \\
Continuous Beta 0 & 0.611699(2) & 4.2(3) \\ 
Continuous Beta 5 & 0.611699(3) & 3.9(4) \\ 
Reverse Monte Carlo 5 & 0.611699(2) & 3.5(2) \\
Reverse Monte Carlo 5 Freeze 100 & 0.611711(4) & 5.1(5) \\
Reverse Monte Carlo 5 Freeze 200 & 0.611704(4) & 5.0(5) \\  
\hline 
\end{tabular} 
}
\end{table}

\begin{table}[!ht]
\caption{Smeared Plaquette 10,000 Trajectories} 
{\centering  
\begin{tabular}{c c c} 
\hline\hline 
Algorithm & Value\ & Integrated Autocorrelation \\ [0.5ex] 
\hline 
Base & 0.9989961(15) & 66(19) \\
Continuous Beta 0 & 0.9989934(8) & 43.(8) \\ 
Continuous Beta 5 & 0.9989935(14) & 58(16) \\ 
Reverse Monte Carlo 5 & 0.9989955(9) & 79(15) \\ 
Reverse Monte Carlo 5 Freeze 100 & 0.9989964(16) & 78(2) \\
Reverse Monte Carlo 5 Freeze 200 & 0.9989973(15) & 64(18) \\  
\hline 
\end{tabular} 
}
\end{table}

\begin{table}[!ht]
\caption{Topological Charge 10,000 Trajectories} 
{\centering  
\begin{tabular}{c c c} 
\hline\hline 
Algorithm & Value\ & Integrated Autocorrelation \\ [0.5ex] 
\hline 
Base & 1.02 $\pm$ 1.05 & 320(140) \\
Continuous Beta 0 & 2.63 $\pm$ 1.06 & 420(180) \\ 
Continuous Beta 5 & 1.80 $\pm$ 1.14 & 260(110) \\ 
Reverse Monte Carlo 5 & -0.23 $\pm$ 0.7 & 190(80) \\ 
Reverse Monte Carlo 5 Freeze 100 & -0.69 $\pm$ 1.05 & 256(110) \\
Reverse Monte Carlo 5 Freeze 200 & 1.23 $\pm$ 1.54 & 530(250) \\ 
\hline 
\end{tabular} 
}
\end{table}
\FloatBarrier
The analysis of the integrated autocorrelation time was carried out with the tools released by the alpha collaboration \cite{Wolff}. Our results so far are largely inconclusive due to low statistics. We have taken a few of the algorithm variants to 100,000 trajectories, however the data is still not precise enough to give any definitive results.
\section{Conclusion}
Although tests so far have not indicated a dramatic improvement of topological freezing, the low statistics of our data prevents us from having firm conclusions. There is a vast parameter space unexplored and more extensive investigation on the dependence of autocorrelation times as functions of the switched coupling and the specifics of the simulation algorithm need to be made. The the path integral interpretation of these algorithms provides a simple method for generalizing these algorithms, possibly allowing us to find an optimal modification that improves the autocorrelation time of topological charge.
We are currently investigating such generalizations that admit action modifications at larger length scales,
comparable to the longest correlation length in the theory. Such an approach may be more likely to produce a favorable outcome closer to the continuum limit. Additionally, we are also extending our statistics. An analysis with about a million trajectories will be available soon.

\begin{acknowledgments}
This project was supported in part by the
U.S.~Department of Energy, Grant Number DE-FG02-04ER41302. KO was also  
supported by the U.S. Department of Energy through Grant Number 
DE-AC05-06OR23177, under which JSA operates the Thomas Jefferson National 
Accelerator Facility. 
\end{acknowledgments}

\end{document}